\documentclass[aps,prl,reprint]{revtex4-1}
\usepackage{amsmath,amssymb,amsfonts}
\usepackage[varg]{txfonts}
\usepackage[]{graphicx}\graphicspath{{graph/}%
{../FEL/"FEL at saturation (with Alberto)"/graph/}}
\usepackage{epstopdf}
\usepackage{hyperref}
\hypersetup{
  colorlinks,
  citecolor=green,
  linkcolor=blue}

\renewcommand{\Re}{{\rm Re\,}}

\begin{document}

\title {Stimulated X-ray Raman scattering in Free Electron Lasers with incoherent spectrum}
\author{
Gennady Stupakov}
\affiliation{SLAC National Accelerator Laboratory, Menlo Park, California 94025, USA}
\author{Max Zolotorev}
\affiliation{Lawrence Berkeley National Laboratory, One Cyclotron Road, Berkeley, California,94720, USA}

\date{\today}

\begin{abstract}
The single-pulse spectrum of self-amplified spontaneous emission (SASE)  free electron lasers (FELs) is characterized by random fluctuations in frequency. The typical spectrum bandwidth for a hard x-ray FEL is in the range of 10-20 eV and is comparable with the distance between  energy levels of valence electrons in atoms an molecules. We calculate the rate of transitions in a quantum three-level system with the energy difference of several eV caused by such radiation and show that for x-ray intensities in the range of $10^{20}$ W/cm$^2$ the probability of the transition over the duration of the x-ray pulse is large. We argue that this effect can be used to modify the spectrum of a SASE FEL potentially making it more narrow. 

\end{abstract} 

\maketitle

X-ray free electron lasers (FELs)~\cite{lcls_2010,AckermannW.:2007qv,Ishikawa:2012ty} can generate highly intense beams of radiation in which nonlinear x-ray--matter interaction plays a dominant role~\cite{Young:2010rz,PhysRevLett.106.083002}. Radiation pulses from FELs of short, sub-femtosecond duration $t$ have the \emph{coherent bandwidth} $\hbar/t$ of several eV which is commensurate with the energy associated with electronic structure of atoms and molecules. Attosecond pulses open a new path for creation of coherent localized valence electronic wave packets for study of the energy transport in pump-probe experiments in molecular systems \cite{PhysRevLett.114.143005,Biggs24092013}.

Generation of subfemtosecond pulses with a large coherent bandwidth  requires operation of an FEL in a special mode~\cite{emma04etal,zholents05-1,PhysRevSTAB.17.120703}, while nominally x-ray pulse duration is in the range of tens, or hundreds, of femtoseconds. Spectra of the typical SASE pulses exhibit fine structure with narrow spikes fluctuating in positions and heights within the relative \emph{incoherent bandwidth} on the order of  $10^{-3}$. This structure is due to the fact that the radiation in a SASE FEL is initiated by the intrinsic shot noise of the electron beam. For the x-ray energy $E_{xr}\approx 10$ keV, the incoherent bandwidth is in the range of 10-20 eV, thus exceeding the distance between the energy levels of valence electrons in atoms in molecules. In this paper, we show that in combination with a high intensity of a focused x-ray beam, through the mechanism of the stimulated Raman scattering,  FEL x-rays can excite quantum levels with the energy distance between them on the order of a few eV. We also argue that the same mechanism leads to ionization of the valence electrons with the cross section that can be many orders of magnitude larger than the direct photoionization cross section for hard x-rays. This effect can play an important role for tightly focused x-ray beams required for single particle imaging~\cite{single_part_imaging}.

\begin{figure}[htb]
\centering
\includegraphics[width=0.3\textwidth, trim=0mm 0mm 0mm 0mm, clip]{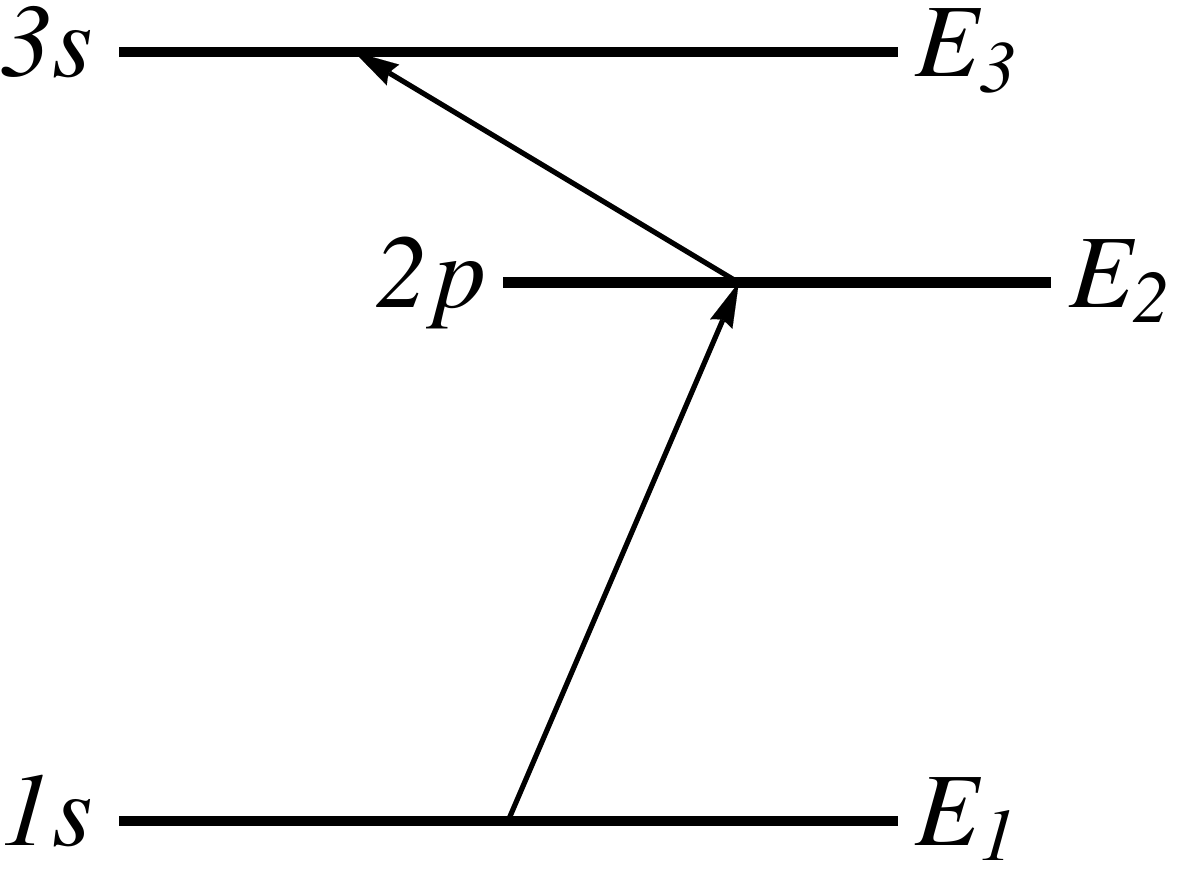}
\caption{Three energy levels in a quantum system. Direct dipole transitions are allowed between levels 1 and 2, and 2 and 3, but not between 1 and 3.}
\label{fig:1}
\end{figure}

We consider a quantum system that has three energy levels $E_1$, $E_2$ and $E_3$, as shown in Fig.~\ref{fig:1}, and assume that there are electric dipole transitions from level 1 to 2 and from 2 to 3, but no direct transitions from 1 to 3 (e.g., $1s$, $2p$ and $3s$  subshells in an atom). 
In the initial state, the system is at the lowest level $E_1$. Stimulated by the incident photons, the system makes a virtual transition to level 2 (which does not require the photon energy to be equal to $E_2-E_1$) and then another virtual transition from 2 to 3. If the spectral width of x-rays is greater that $E_3-E_1$, the transition probability $1\to 3$ is proportional to the square of the x-ray spectral power density and becomes large for intensities that can be achieved in a focussed FEL beam.  

To calculate the transition probability $1\to 3$ we treat the x-ray radiation in the semiclassical approximation~\cite{scully1997quantum}. The linearly polarized electric field is written as ${\cal E}(t)\cos(\omega_0 t)$, where $\omega_0 = E_{xr}/\hbar$ is the central frequency and ${\cal E}(t)$ is a slow varying amplitude of the incident FEL radiation. Variation of ${\cal E}(t)$ in time determines the width of the x-ray spectrum; because of the stochastic nature of the SASE radiation we  treat ${\cal E}(t)$ as a stationary random process with a given statistical properties.

We start from the equations for the time evolution of the probability amplitudes $a_1$, $a_2$ and $a_3$ for the corresponding energy levels~\cite{scully1997quantum}: 
    \begin{subequations}\label{eq:1}
    \begin{align}
    \label{eq_c1}
    \dot a_1
    &=
    \epsilon 
    V(t) a_2 \cos
    \left(
    \omega_0 t
    \right)\,
    ,\\
    \label{eq_c12}
    \dot a_2
    &=
    -
    \epsilon
    V^*(t) a_1 \cos
    \left(
    \omega_0 t
    \right)
    +
    \epsilon 
    U(t) a_3 \cos
    \left(
    \omega_0 t
    \right)\,
    ,\\
    \label{eq_c3}
    \dot a_3
    &=
    -
    \epsilon 
    U^*(t) a_2 \cos
    \left(
    \omega_0 t
    \right)\,,
    \end{align}
    \end{subequations}
where the matrix elements $U(t)$ and $V(t)$ are given by
    \begin{align}\label{eq:2}
    V(t)
    &=
    -\frac{i}{\hbar}
    \langle 1|H_{int}|2\rangle
    e^{-i(E_2-E_1)t/\hbar}
    =
    -\frac{i}{\hbar}
    d_{12} {\cal E}(t)
    e^{-i \omega_{12}t}
    \,,
    \nonumber\\
    U(t)
    &=
    -\frac{i}{\hbar}
    \langle 2|H_{int}|3\rangle
    e^{-i(E_3-E_2)t/\hbar}
    =
    -\frac{i}{\hbar}
    d_{23} {\cal E}(t)
    e^{-i \omega_{23}t}
    \,,
    \end{align}
with $H_{int}$ the interaction Hamiltonian in the dipole approximation, and $d_{12}$ and $d_{23}$ the matrix elements of the dipole operator between the corresponding levels. We assume that the time evolution of the amplitudes $a_i$ is slow in comparison with $\omega_0^{-1}$; this is indicated by the formal small parameter $\epsilon$ in Eqs.~\eqref{eq:1}:
    \begin{align}\label{eq:5}
    \epsilon
    \sim
    \frac{{\cal E}d}{\hbar\omega_0}
    \ll
    1,
    \end{align}
where $d$ is the characteristic dipole matrix element involved into the transitions. Eq.~\eqref{eq:5} states that the Rabi frequency ${\cal E}d/\hbar$ is much smaller than the x-ray frequency $\omega_0$. Note that Eqs.~\eqref{eq:1} conserve the total probability $|a_1|^2 + |a_2|^2 + |a_3|^2$.

Using the smallness of $\epsilon$ we now average Eqs.~\eqref{eq:1} over the rapid oscillations with frequency $\omega_0$, and obtain simplified equations for a slow variation of amplitudes $a_i$ on a time interval much larger than $1/\omega_0$. The starting point for this approximation is the following representation of the amplitudes:
    \begin{align}\label{eq:3}
    a_i
    \approx
    \alpha_i(\epsilon t)
    +
    \epsilon
    \beta_i(\epsilon t)
    \sin
    \left(
    \omega_0 t
    \right)
    +
    \epsilon^2
    \gamma_i(\epsilon t)
    \cos
    \left(
    \omega_0 t
    \right)
    +
    \ldots
    \end{align}
where $\alpha_i$, $\beta_i$ and $\gamma_i$ are slow varying functions of time which is indicated by their argument being $\epsilon t$. Strictly speaking, on the right-hand side,  there should also be terms with harmonics of the frequency $\omega_0$, however, they do not contribute to the final result, and are neglected. Note that $\beta_i$ and $\gamma_i$ are small corrections to the zeroth order amplitudes $\alpha_i$, which are the subjects of our interest. Substituting Eq.~(\ref{eq:3}) into~(\ref{eq_c1})  and collecting terms in front of $\cos(\omega_0 t)$ and $\sin(\omega_0 t)$ yields
    \begin{align*}
    \beta_1
    =
    \frac{V\alpha_2}{\omega_0}
    ,\qquad
    \gamma_1
    =
    \frac{
    \dot
    \beta_1}{\omega_0}
    \,.
    \end{align*}    
Averaging Eq.~(\ref{eq_c1}) over the fast period $2\pi/\omega_0$ shows that the rate of change of $\alpha_1$ is of the second order
    \begin{align*}
    \dot
    \alpha_1
    =
    \frac{1}{2}
    \epsilon^2
    V
    \gamma_2
    \,.
    \end{align*}    
Repeating the same analysis for the second and third equations in~\eqref{eq:1} we arrive at the following set of differential equations for the slow varying parts of the amplitudes:
    \begin{subequations}\label{eq:4}
    \begin{align}    
    \label{eq:4-1}
    \dot
    \alpha_1
    &=
    \epsilon^2
    \frac{1}{2\omega_0^2}
    V
    \frac{d}{dt}
    \left(
    -
    V^*\alpha_1
    +
    U\alpha_3
    \right)\,,
     \\
    \label{eq:4-2}
    \dot
    \alpha_2
    &=
    -
    \epsilon^2
    \frac{1}{2\omega_0^2}
    \left(
    V^*
    \frac{d}{dt}
    V\alpha_2
    +
    U
    \frac{d}{dt}
    U^*\alpha_2
    \right)\,,
     \\
    \label{eq:4-3}
    \dot
    \alpha_3
    &=
    \epsilon^2
    \frac{1}{2\omega_0^2}
    U^*
    \frac{d}{dt}
    \left(
    V^*\alpha_1
    -
    U\alpha_3
    \right)\,.
    \end{align}
    \end{subequations}   
Note that the second equation is decoupled from the first and the third ones; it will be omitted from the subsequent analysis.  

The system of equations~\eqref{eq:4} conserves the probability $|a_1|^2 + |a_3|^2$ only approximately. It is easy to derive from Eqs.~\eqref{eq:4-1} and~\eqref{eq:4-2} that the following combination remains constant,
    \begin{align*}
    |\alpha_1|^2
    +
    |\alpha_3|^2
    +
    \frac{\epsilon^2}{2\omega_0^2}
    \left[
    |\alpha_1|^2 |V|^2
    +
    |\alpha_3|^2 |U|^2
    -
    2
    \Re
    \left(U\alpha_3
    \alpha_1^*V
    \right)
    \right]
    ,
    \end{align*}
which differs from the sum of probabilities $|a_1|^2 + |a_3|^2$ by the last term. This term is small for $\epsilon\ll 1$. We will drop the formal smallness parameter $\epsilon$ in what follows.

We now use the initial condition that at time $t=0$ the system is at the lowest energy level $E_1$, $\alpha_1(t=0) = 1$ and $\alpha_3(t=0) = 0$. Considering  time intervals small enough that the probability to find the system at level 3  remains small, we have $|\alpha_3(t)|\ll 1$, and $\alpha_1(t) \approx 1$. To the lowest order, we substitute $\alpha_1 = 1$ and $\alpha_3  = 0$ into the right-hand side of Eq.~\eqref{eq:4-3} to obtain
    \begin{align*}
    \dot
    \alpha_3
    &=
    -
    \frac{1}{2\hbar^2\omega_0^2}
     d_{12} d_{23}
    {\cal E}(t)
    e^{i \omega_{23}t}
    \frac{d}{dt}
    {\cal E}(t)
    e^{i \omega_{12}t}
    \,,
    \end{align*}
where we have also used Eqs.~\eqref{eq:2} for the matrix elements. Integrating this equation over time and calculating the probability to find the system at level 3 at time $t$ gives
    \begin{align}\label{eq:6}
    & w_{3}
    \equiv
    |\alpha_3(t)|^2
    =
    \frac{1}{4\hbar^4\omega_0^4}
     d_{12}^2 d_{23}^2
    \int_0^t
    \int_0^t
    dt'dt''
    {\cal E}(t'')
    e^{-i \omega_{23}t''}
    \nonumber\\
    &\times
    {\cal E}(t')
    e^{i \omega_{23}t'}
    \frac{d}{dt''}
    \left(
    {\cal E}(t'')
    e^{-i \omega_{12}t''}
    \right)
    \left(
    \frac{d}{dt'}
    {\cal E}(t')
    e^{i \omega_{12}t'}
    \right)
    \,.
    \end{align}

We now assume that the electric field ${\cal E}(t)$ is a stationary random function which, after decomposition into the Fourier integral,
    $
    {\cal E}(t)
    =
    \int_{-\infty}^{\infty}
    d\omega\,
    \hat {\cal E}(\omega)
    e^{-i\omega t}
    $,
can be characterized by the correlator
    \begin{align}\label{eq:7}
    \langle
    \hat {\cal E}(\omega)
    \hat {\cal E}(\omega')
    \rangle
    =
    W(\omega)
    \delta(\omega+\omega')
    \,,
    \end{align}
where the brackets denote an ensemble averaging and $W(\omega)$ is the spectrum of the electric field measured relative to the central frequency $\omega_0$. To carry out the statistical averaging of the probability $w_{3}$, one has to substitute the Fourier representation for ${\cal E}(t)$ into~\eqref{eq:6} and calculate the fourth order correlators $\langle\hat{\cal E}(\omega_1)\hat{\cal E}(\omega_2)\hat{\cal E}(\omega_3)\hat{\cal E}(\omega_4)\rangle$. With an additional assumption that ${\cal E}(t)$ is a Gaussian random process, these correlators are expressed as a sum of the products of the second order correlators~\cite{goodman1985statistical} for which we can use Eq.~\eqref{eq:7}. After a straightforward calculation one finds
    \begin{align}\label{eq:8}
    w_{3}
    &=
    \frac{16\pi^3d_{12}^2 d_{23}^2}{c^2\hbar^4\omega_0^4}t
    \left(
    \omega_{23}
    -
    \omega_{12}
    \right)^2
    \int_{0}^{\infty}
    d\omega
    P\left(\omega-\omega_{13}\right)
    P\left(\omega\right)
    \,,
    \end{align}
where instead of the spectral function $W(\omega)$ we now use  the spectral power of the x-ray radiation $P(\omega)$, $P(\omega)=({c}/{8\pi})W(\omega)$. It is important to emphasize here that $P(\omega)$ is the FEL spectrum averaged over many pulses; while a single-pulse SASE spectrum exhibits many spikes, the average one is a smooth function of  frequency.

Note that the probability~\eqref{eq:8} vanishes if level 2 is in the middle between levels 1 and 3, that is $\omega_{13}=\omega_{23}$. This is a well known effect of vanishing Raman scattering in three-level system(see, e.g., \cite{budker2008atomic}, p. 185).

To illustrate the feasibility of the stimulated Raman scattering for typical FEL parameters, we will now estimate the probability of transitions in a hydrogen atom from level $1s$ (level 1) to level $3s$ (level 3) through level $2p$ (level 2). Measuring the energy from the lowest level, we have $E_1=0$, $E_2=\frac{3}{4}\mathrm{Ry}$ and $E_3=\frac{8}{9}\mathrm{Ry}$, and for the dipole moments, $d_{12}=2^{15/2}3^{-9/2}ea_B$, $d_{23}=3^{4/3}ea_B$, where $a_B$ is the Bohr radius~\cite{landau2013quantum}. We take the FEL parameters close to the ones in the experiment~\cite{Fuchs:2015fk}, assuming the  pulse energy of 1.0 mJ, the pulse duration of 50 fs with a flat temporal pulse profile, and $\hbar\omega_0= 10$ keV. The beam is focused onto 150 nm$\times$150 nm spot size with the intensity $P_0=1.2\times 10^{20}$ W/cm$^2$. To simplify calculations, we take for the averaged x-ray spectrum a Gaussian profile with the rms spread $\Delta\omega\approx 2\times10^{-3}\omega_0=20\ \mathrm{eV}/\hbar$,
    $
    P(\omega)
    =
    (2\pi)^{-1/2}\Delta\omega^{-1}P_0
    e^{-\omega^2/2\Delta\omega^2}
    $
(we remind the reader that the frequency $\omega$ in this equation is measured relative to the central frequency $\omega_0$). Carrying out the integration in Eq.~\eqref{eq:8} we find for the probability $w_3$:
    \begin{align}\label{eq:9}
    w_{3}
    &=
    t
    P_0^2
    \frac{8\pi^{5/2}d_{12}^2 d_{23}^2}{c^2\hbar^4\omega_0^4\Delta\omega}
    \left(
    \omega_{23}
    -
    \omega_{12}
    \right)^2
    e^{-\omega_{13}^2/4\Delta\omega^2}
    ,
    \end{align}
which for our example gives for the transition probability 
    \begin{align}\label{eq:10}
    w_{3}
    \approx
    0.021
    t\ \mathrm{[fs]}
    .
    \end{align}
Since our calculations assume $w_{3} \ll 1$, this formula is valid for $t\lesssim 20$ fs. Note that the smallness parameter~\eqref{eq:5} estimated with ${\cal E}\sim  \sqrt{8\pi P_0/c}\approx 3\times 10^{13}$ V/m and $d\sim \sqrt{d_{12}d_{23}} =  2.4 ea_B$ equals 0.37 which is not small compared to unity. 

To test the accuracy of our approximate analysis we numerically integrated Eqs.~\eqref{eq:1} 
\begin{figure}[htb]
\centering
\includegraphics[width=0.4\textwidth, trim=0mm 0mm 0mm 0mm, clip]{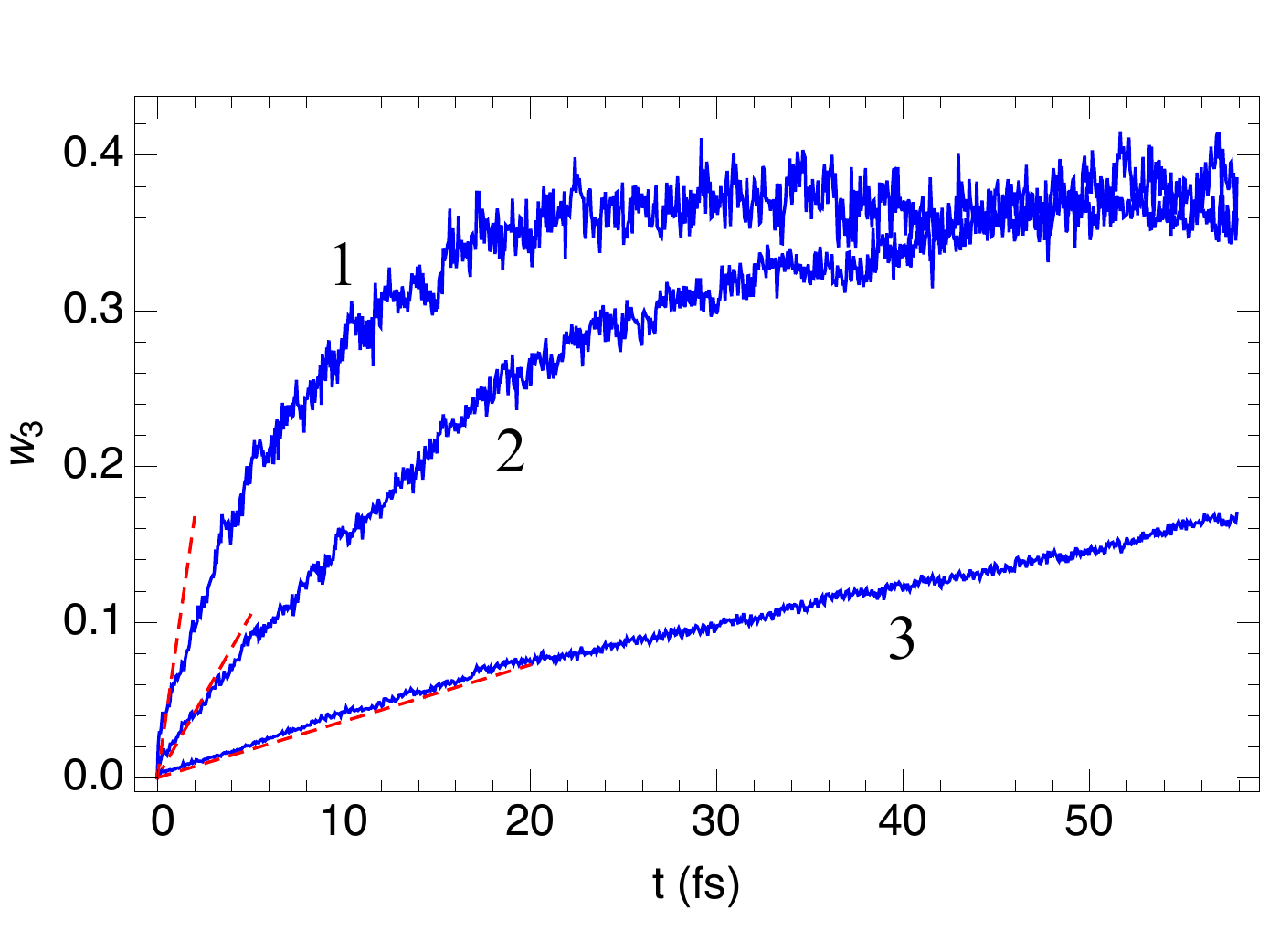}
\caption{Time evolution of the probability $w_3$ averaged over 200 realizations of the random field ${\cal E}(t)$: 1)-$P_0=2.4\times 10^{20}\ \mathrm{W/cm}^2$, 2)-$P_0=1.2\times 10^{20}\ \mathrm{W/cm}^2$ and 3)-$P_0=5\times 10^{19}\ \mathrm{W/cm}^2$. The red dashed lines show the small-time approximation~\eqref{eq:9} for each case.}
\label{fig:2}
\end{figure}
for 200 realizations of the random electric field ${\cal E}(t)$ with a Gaussian spectrum  described above, for three different intensities:  $P_0=2.4\times 10^{20}\ \mathrm{W/cm}^2$, $P_0=1.2\times 10^{20}\ \mathrm{W/cm}^2$ and $P_0=5\times 10^{19}\ \mathrm{W/cm}^2$. The plots of $w_3(t)$ as a function  of time for the three cases are shown in Fig.~\ref{fig:2}.
The red dashed lines near the origin show the small-time approximation calculated with Eq.~\eqref{eq:9} for each case. Remarkably, even though the parameter~\eqref{eq:5} is not really small for cases 1 and 2, Eq.~\eqref{eq:9} gives a relatively good approximation for the initial slope of $w_3(t)$. One can also see that, for these two cases, after an initial, approximately linear, growth $w_3(t)$ saturates at the level $w_3(t)\approx 0.4$. 

From Eq.~\eqref{eq:8} we can derive the cross section for the scattering  replacing $P(\omega)$ in~\eqref{eq:8} by $c\hbar\omega n_{ph}(\omega)$ where $n_{ph}(\omega)$ is the density of photons in the beam per unit frequency interval. We then re-write Eq.~\eqref{eq:8} as an expression for the probability per unit time
    \begin{align}\label{eq:11}
    \frac{w_{3}}{t}
    &=
    c
    \int_{0}^{\infty}
    d\omega\,
    \sigma(\omega)
    n_{ph}(\omega)
    \,,
    \end{align}
where $\sigma(\omega)$ has a meaning of the differential cross section for the scattering,
    \begin{align}\label{eq:12}
    \sigma(\omega)
    =
    \frac{16\pi^3d_{12}^2 d_{23}^2}{c^2\hbar^3\omega_0^3}
    (
    \omega_{23}
    -
    \omega_{12}
    )^2
    P(\omega-\omega_{13})
    .
    \end{align}
Note that this cross section is proportional to the incident intensity of x-rays at the frequency shifted by the distance between the level 1 and 3. For our numerical example~\eqref{eq:10}, the maximum cross section at $\omega=\omega_{13}$, is $\sigma\approx 5\times 10^{-22}\ \mathrm{cm}^{2}$. This cross section is almost five orders of magnitude larger than the ionization cross section of hydrogen   by 10 keV photons, $\sigma_\mathrm{ion}\approx 9\times 10^{-27}\ \mathrm{cm}^{2}$, and three orders of magnitude larger than the Thomson cross section for elastic scattering.

\begin{figure}[htb]
\centering
\includegraphics[width=0.4\textwidth, trim=0mm 45mm 0mm 50mm, clip]{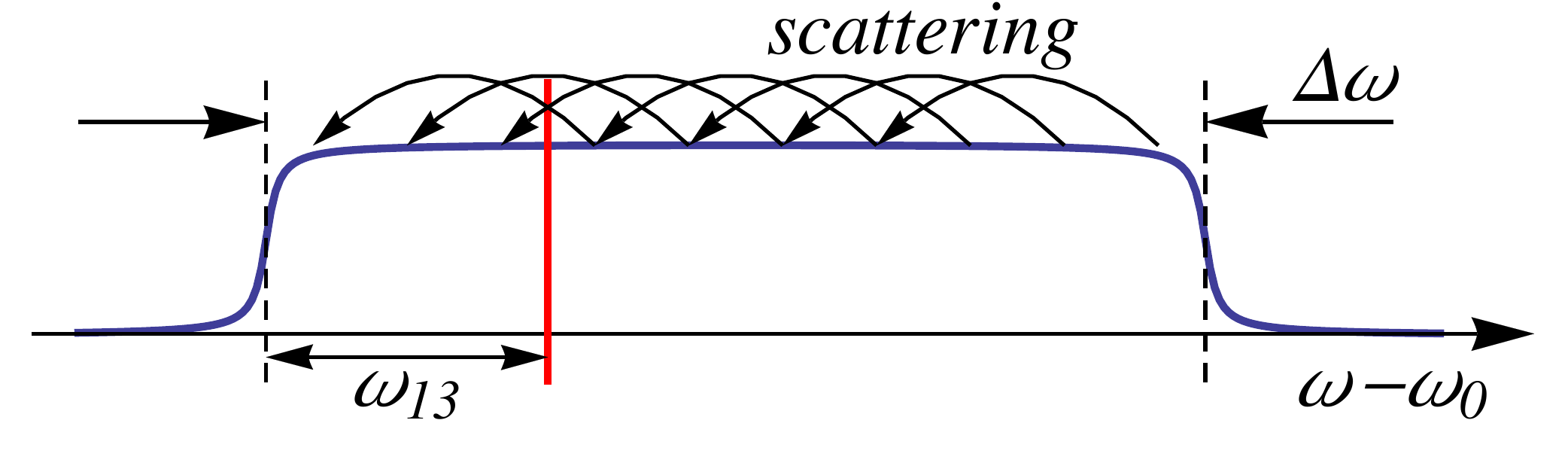}
\caption{Illustration of the x-ray spectrum evolving due to the stimulated Raman scattering. The scattering processes occur only to the right of the red vertical line located at the distance $\omega_{13}$ from the left edge of the spectrum. The scattering downshifts photons in frequency as indicated by the curved arrows.}
\label{fig:3}
\end{figure}

Examination of Eq.~\eqref{eq:12} shows that for radiation with a bandwidth smaller than $\omega_{13}$ the cross section vanishes because $P(\omega-\omega_{13})$ lies outside of the bandwidth if $\omega$ is inside it. For a spectrum wider than $\omega_{13}$ the scattering is  different at the low-energy and high-energy parts of the spectrum. Indeed, assuming for illustration a flat spectrum occupying the interval $[\omega_1, \omega_2]$ of width $\Delta\omega = \omega_2-\omega_1>\omega_{13}$ (see Fig.~\ref{fig:3}) we see that the scattering occurs only  in the region $[\omega_1+\omega_{13},\omega_2]$, while at the low-energy end of the spectrum $[\omega_1, \omega_1+\omega_{13}]$ the cross section~\eqref{eq:12} is zero. Taking into account that in an act of scattering a photon of frequency $\omega$ changes its frequency to $\omega-\omega_{13}$, we expect that, given enough scattering events,  the stimulated Raman scattering would lead to a noticeable modification, and possible shrinking, of the incident spectrum of x-rays. 
For a numerical example, let us assume that x-rays are passing through a frozen solid hydrogen with the density $5.4\times 10^{22}\ \mathrm{atom/cm}^{3}$. For the cross section $\sigma\approx 5\times 10^{-22}\ \mathrm{cm}^{2}$ estimated above, one needs the hydrogen target thickness of $\approx 0.4$ mm for every photon in the beam to experience a scattering event during the passage through the medium. Note that at this distance the divergence of the x-ray beam focussed onto the focal spot size of $\sim 100$ nm can be neglected.

While the above analysis indicates the feasibility to modify the spectrum of the x-rays through the mechanism of stimulated Raman scattering, a more accurate, quantum treatment of the problem is needed to be able to draw quantitative conclusions about the effect.  

To elucidate the underlying physical mechanism, in our analysis above, we have considered a model of a three-level quantum system. In reality, in atoms and molecules, the stimulated scattering would cause multi-level transitions occurring at the same time with different frequencies and at various rates. Our numerical results should then be considered as a  guide only; a more accurate analysis is required of the quantum dynamics in a multi-level system interacting with stochastic incident field. We would also point out, that it seems highly plausible that the same mechanism of the stimulated Raman scattering will lead to electron transitions into continuum part of the spectrum, effectively ionizing atoms and molecules with the cross section much larger that the direct photoionization by x-rays. 

In conclusion, we have shown that, in a three-level system, tightly focussed SASE FEL radiation can lead to excitation, and likely ionization, of valence electrons in atoms and molecules through the mechanism of the stimulated Raman scattering. The cross section for the scattering can be large enough to be used for modification of the FEL spectrum by sending the x-ray beam through a medium with properly selected energy levels, thus opening up an opportunity to modify and control the SASE spectrum before it is used in an experiment.

The authors would like to thank P. Bucksbaum, D. Budker and J. Hastings for stimulating discussions. G.~S. acknowledges support from the DOE grant No. DE-AC02-76SF00515, and M.~Z. acknowledges support from DOE grant No. DE-AC02-05CH11231. 

%

\end{document}